\documentclass[twocolumn,prd,showpacs,amsmath,amssymb]{revtex4-1}

\usepackage{graphicx}% Include figure files
\usepackage{dcolumn}% Align table columns on decimal point
\usepackage{bm}% bold math
\usepackage[cp866]{inputenc}
\begin{document}
\title{Photon-axion mixing
and ultra-high-energy cosmic rays from BL Lac type objects
--- Shining light through the Universe.}
\author{M.~Fairbairn}
\affiliation{Physics, King's College London, Strand, WC2R 2LS, London, UK}
\author{T.~Rashba}
\affiliation{Max-Planck-Institut f\"ur Sonnensystemforschung,
Max-Planck-Str. 2, D-37191 Katlenburg-Lindau, Germany;}
\affiliation{Pushkov Institute of Terrestrial Magnetism, Ionosphere and
Radio Wave Propogation
of the Russian Academy of Sciences, 142190, Troitsk, Moscow region, Russia}
\author{S.~Troitsky}
\affiliation{Institute for Nuclear Research of the Russian Academy of
Sciences,
60th October Anniversary Prospect 7a, 117312, Moscow, Russia}
\date{}

\begin{abstract}
Photons may convert into axion like particles and back in the magnetic
field of various astrophysical objects, including active galaxies,
clusters of galaxies, intergalactic space and the Milky Way.  This is a
potential explanation for the candidate neutral ultra-high-energy
($E>10^{18}$~eV) particles from distant BL Lac type objects which have
been observed by the High Resolution Fly's Eye experiment. Axions of the
same mass and coupling may explain also TeV photons detected from distant
blazars.
\end{abstract}
\pacs{14.80 Va, 98.70 Sa}
\maketitle

\section{Introduction}
\label{sec:intro}

Axions are pseudo-scalar particles which arise as the Nambu-Goldstone
Bosons of the broken Peccei-Quinn symmetry \cite{pq}.  They obtain a mass
when the CP violating QCD theta term is driven to zero in agreement with
observations \cite{wil, wein}.  When motivated in this way, the
relationship between the axion mass and coupling is related to the pion
mass and decay constant such that for a given axion mass, the coupling to
photons is determined up to factors of order a few ($m M \sim m_\pi f_\pi$
where $m$ is the axion mass and $M$ is the inverse axion coupling, see
section \ref{sec:BLL:axion:in-source}).  While considerable experimental
and theoretical work has eliminated much of the parameter space of such
models, axions are still a viable candidate for both the solution of the
strong-CP problem and for cold dark matter.

The term Axion Like Particle refers to a particle with a similar
Lagrangian structure to the Peccei-Quinn axion but where the constraints
on the parameters of the Lagrangian have been relaxed.  In other words
there may be particles like the axion weakly coupled to the Standard Model
even if they do not solve the strong-CP problem.  For simplicity, in the
rest of this paper, we shall refer to all such particles as axions, while
the particular kind of particle associated with the solution of the
strong-CP problem we shall refer to as the Peccei-Quinn axion.

Axions have been invoked to solve a variety of different problems in
physics and astrophysics.  For example, it has suggested that they might
be responsible for the dimming of supernovae -- photons from distant type
Ia supernovae might convert into axions as they cross the Universe to
reach us which may explain the apparent low luminosity of high redshift
supernovae normally subscribed to the presence of a cosmological constant
\cite{ckt}.  Such models are interesting, although there may be problems
with the frequency dependence of the dimming effect and they appear
difficult to reconcile with baryon acoustic oscillation observations
\cite{goob, fairb, bao}.

In Ref.~\cite{Csaki}, photon-axion oscillations in intergalactic space
have
been suggested as an explanation of super-GZK cosmic rays detected by the AGASA
experiment although mixing in the source was not considered. Since such mixing means
that photons spend some of their time as axions while on route to earth, the
attenuation length of photons is effectively increased.
Unfortunately it seems that for the parameters of Ref.
\cite{Csaki}, the original flux of photons in the source
should exceed the flux observed at the Earth by several orders of
magnitude - all these additional photons have to lose their energy in
cascades on the background radiation which would be in conflict with
EGRET and FERMI limits on the diffuse gamma-ray background.

More recently, the detection of TeV photons from objects at cosmological
distances has led to a reconsideration of axions.  It is difficult to
explain how such photons could reach the Earth given the opacity of the
Universe at those wavelengths due to pair production on the background
infrared radiation.  It has been suggested that the mixing of photons with
axions in the intergalactic magnetic field may explain this, although the
required intergalactic magnetic field has to be on the high side
\cite{roncadelli} (for a more recent work and review, see
Ref.~\cite{DARMA2011}).
Another suggestion is
that
photons are converted into axions in the magnetic field of the active
galaxy itself, which is a rather reasonable assumption for axions with low
masses.  If such a mixing were to take place efficiently, up to one third
of the initial high energy photon flux may cross the Universe in the form
of axions before being converted back into photons in the magnetic field
of the Milky Way, avoiding the attenuation that photons would experience
as they travel across the Universe.  The authors of \cite{serpico}
identified the axion parameters and galactic magnetic field which can
explain the arrival of TeV photons from cosmological sources. In this
note, we shall analyse these axion scenarios to see if
they might also explain the origin of apparently neutrally charged
ultra high energy cosmic rays which may come from distant extragalactic
sources -- BL Lac type objects~\cite{GTTT:HiRes, HiRes:BL}.

One of the most fascinating predictions of theories which contain axions
is the idea that one may 'shine light through walls' by converting photons
to and from axions on either side of a wall using strategically placed
magnetic fields.  In this work we are doing the same experiment but we are
using the Universe as our wall and galaxies and their environment for our
magnetic fields.

In section \ref{sec:BLL:axion:in-source} we shall go over the mathematics
of the mixing phenomenon and discuss the mixing of photons and axions in
astrophysical sources.  Then in section \ref{sec:BLL:axion:MW} we will
discuss the evidence for the arrival of ultra high energy cosmic rays from
directions coincident with BL Lac objects before discussing photon-axion
mixing as a possible explanation for these events in section \ref{expl}.
%We will then discuss the magnetic field of the Milky Way that we will
%assume to perform our analysis in section \ref{sec:mixGMF} before
%preforming the search for correlations between regions of large B-field
%and cosmic ray events in section \ref{sec:results}.
Finally we will list
some of the consequences of this model and other ways to test it before
moving to our conclusions.

\section{Photon-axion mixing in astrophysical objects.}
\label{sec:BLL:axion:in-source}
The Lagrangian describing the photon and axion takes the following form
(similar results hold for a scalar),
\[
\mathcal{L}=\frac{1}{2}(\partial^\mu a\partial_\mu a-m^2 a^2)
-\frac{1}{4}\frac{a}{M}F_{\mu\nu}\widetilde
F^{\mu\nu}-\frac{1}{4}F_{\mu\nu}F^{\mu\nu},
\]
where $F_{\mu\nu}$ is the electromagnetic stress tensor and
$\widetilde
F_{\mu\nu}=(1/2)\epsilon _{\mu \nu \rho \lambda }F_{\rho \lambda }$
is its dual,
$a$ denotes the pseudo-scalar axion, $m$ is the axion mass and $M$ is the
inverse axion-photon coupling. Because of the
$F_{\mu\nu}\tilde{F}^{\mu\nu}$ term, there is a finite probability for the
photon to mix with the axion in the presence of a magnetic field.  Mixing
also occurs between photon components with different polarizations
\cite{sikivie, raffelt}. We will be interested in light, $m
\lesssim 10^{-5}$~eV, axions with inverse coupling mass scale $M\sim$
few$\times 10^{10}$~GeV.
% which are the parameters that can explain the
%observations of TeV photons from cosmological sources.
For
axions of these masses the most stringent bound on the coupling,
$M>1.1\times 10^{10}$~GeV at the 95\% CL, has been placed by the CAST
experiment~\cite{CAST2008}.

Technically, the mixing may be described as follows. We represent the photon
field $A(t,x)$ as a superposition of fixed-energy components
$A(x)e^{-i\omega t}$. If the magnetic field does not
change significantly on the photon wavelength scale and the index of refraction of the medium
$|n-1|\ll 1$, one can decompose~\cite{raffelt} the operators in the field
equations as (for a photon moving in the $z$ direction)
$\omega^2+\partial_z^2\rightarrow
2\omega(\omega-i\partial_z)$, so that the field equations become
Schrodinger-like,
\begin{equation}
i\partial_z \Psi=
-\left(\omega+\mathcal{M}\right)\Psi\qquad;\qquad
\Psi=\left(\begin{array}{c}A_{x}\\A_{y}\\a\end{array}\right),
\label{schrodinger}
\end{equation}
where
\[
%\label{mixmat}
{\cal M}\equiv\left(
\begin{array}{ccccccccc}
\Delta_p+\Delta_{{\rm Q},\parallel}&0&\Delta_{Mx}\\
0&\Delta_p+\Delta_{{\rm Q},\perp}&\Delta_{My}\\
\Delta_{Mx}&\Delta_{My}&\Delta_m
\end{array}
\right).\hspace{0.7cm}
\]
%\end{equation}
The mixing is determined by the refraction parameter $\Delta_p$,
the axion-mass parameter $\Delta_m$, the mixing parameter $\Delta_M$
and the QED dispersion parameter $\Delta_{{\rm Q},\perp}$. The first three
parameters
are equal to
\[
\begin{array}{rcccl}
\displaystyle \Delta_{M\! i} \!\!\!&=&\!\!\! \displaystyle \frac{B_i}{2M}
\!\!\!&=&\!\!\! \displaystyle
153
\left(\frac{B_i}{1~\mbox{G}}\right)
\left(\frac{10^{10}~\mbox{GeV}}{M}\right)\mbox{pc}^{-1}\!,\\
\displaystyle \Delta_m \!\!\!&=&\!\!\! \displaystyle \frac{m^2}{2
\omega} \!\!\!&=&\!\!\! \displaystyle 7.8\times
10^{-11}\left(\frac{m}{10^{-7}~\mbox{eV}}\right)^2
\left(\frac{10^{19}~\mbox{eV}}{\omega}\right)\mbox{pc}^{-1}\!,\\
\displaystyle \Delta_p \!\!\!&=&\!\!\! \displaystyle \frac{\omega
_p^2}{2 \omega } \!\!\!&=&\!\!\! \displaystyle 1.1\times
10^{-6}\!\!\left(\frac{n_e}{10^{11}~\mbox{cm}^{-3}}\right)
\!\left(\frac{10^{19}~\mbox{eV}}{\omega}\right)\mbox{pc}^{-1}\!,
\end{array}
\]
respectively.  Here $\omega _p^2=4\pi \alpha n_e/m_e$ is the plasma
frequency squared (effective photon mass squared), $n_e$ is the electron
density, $B_i$, $i=x,y$ are the the components of the magnetic field $B$,
~ $m_e$ is the electron mass, $\alpha $ is the fine-structure constant and
$\omega $ is the photon (axion) energy.

\begin{figure}
\begin{center}
\includegraphics[width=0.9\columnwidth]{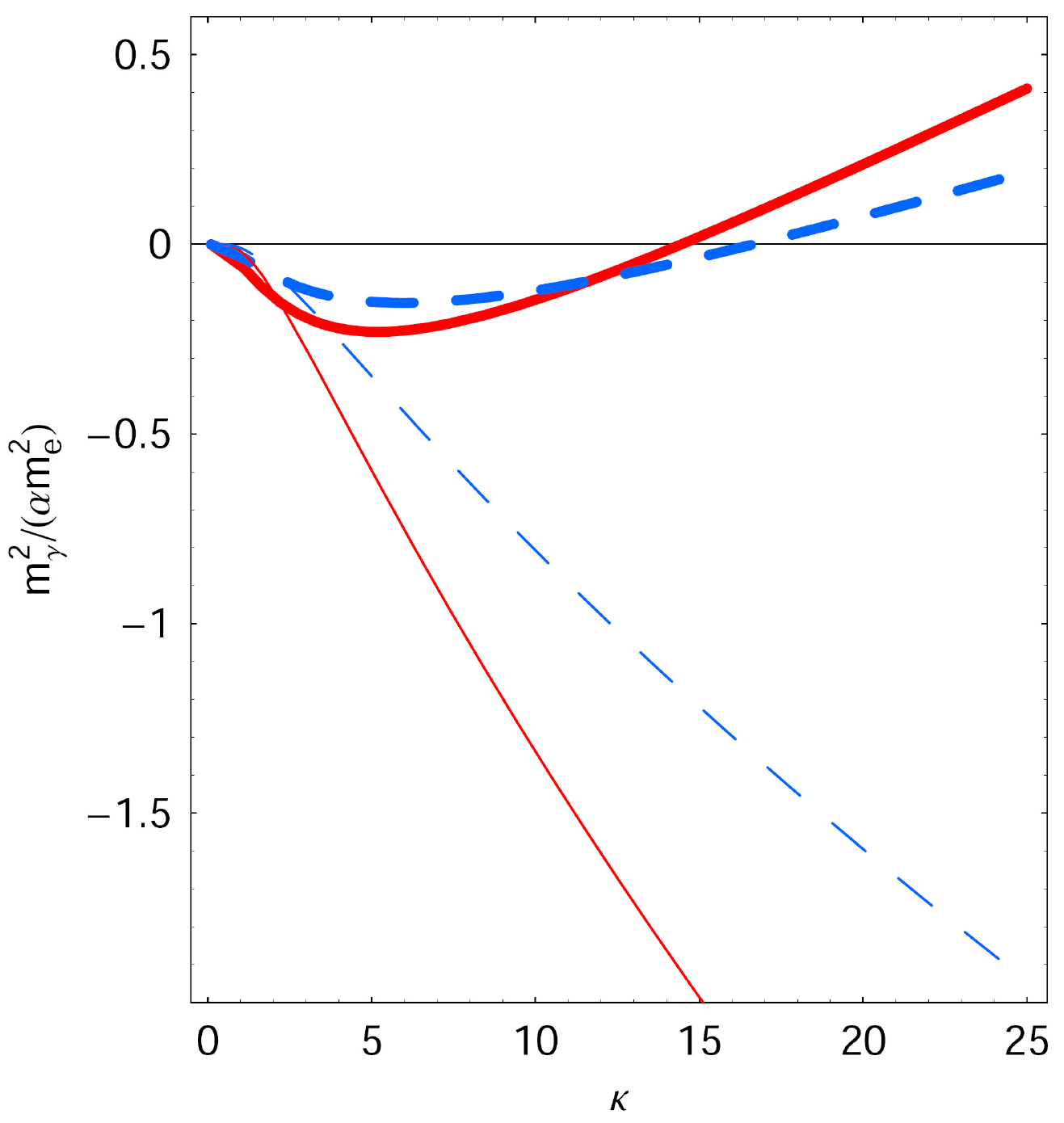}
\caption{
\label{fig:ritus}
Effective mass squared~\cite{Ritus} of the transverse (red full lines) and
longuitudinal (blue dashed lines) photon in the external magnetic field,
expressed in terms of $\alpha m_{e}^{2}$, as a function of $\kappa =
(\omega/(m_{e}))(B/B_{\rm cr})$. Thick lines represent the real part and
thin lines represent the imaginary part.}
\end{center}
\end{figure}

The QED dispersion parameter is
\[
\Delta_{{\rm Q},\parallel(\perp)}=\frac{m^{2}_{\gamma
,\parallel(\perp)}}{2\omega},
\]
where $m^{2}_{\gamma ,\parallel(\perp)}$ is the effective mass square of
the longuitudinal (transverse) photon which arises due to interaction with
the external magnetic field. This quantity has been calculated in
Ref.~\cite{Ritus} (see also Ref.~\cite{Toll} for a similar but less
explicit result),
\begin{equation}
m^{2}_{\gamma ,\parallel(\perp)}  =
\frac{\alpha m_{e}^{2}}{6\pi}\int\limits_{1}^{\infty}\!du\,
\frac{8u+1 \mp 3}{z u \sqrt{u(u-1)}}f'(z),
\label{photon-mass}
\end{equation}
where
\[
z=\left(\frac{4u}{\kappa}   \right)^{2/3}
\]
and
\begin{equation}
\kappa=\frac{1}{m_{e}^{3}} \sqrt{(eF_{\mu\nu}l_{\nu})^{2}}=
\frac{\omega}{m_{e}} \, \frac{B_{\perp}}{B_{\rm cr}}\approx
0.44 \left(\frac{\omega}{10^{19}~\mbox{eV}}\right) \left(\frac{B_{\perp}}{1~{\rm
G}} \right)
\label{Eq:kappa}
\end{equation}
$F_{\mu\nu}$ is the electromagnetic
stress tensor, $l_{\nu }$ is the photon 4-momentum, $B_{\perp}$ is the
component of the magnetic field perpendicular to the photon propagation and
$B_{\rm cr}=m_{e}^{2}/e\approx 4.4 \times 10^{13}$~G;
\[
f(z)=i \int\limits_{0}^{\infty}\!dt\,{\rm e}^{-i(zt+t^{3}/3)}
\]
and the real and imaginary parts of the function $f(z)$ may be expressed
explicitly through the Airy functions.
We plot the real and imaginary parts
of the squared mass of the longuitudinal and transverse photons in
Fig.~\ref{fig:ritus} which is similar to Fig.~1 of Ref.~\cite{Ritus}.
In the region $\kappa\ll 1$, which is often quoted,
Eq.~(\ref{photon-mass}) may be approximated as follows,
\begin{equation}
m^{2}_{\gamma ,\parallel(\perp)}
\approx
\alpha m_{e}^{2}
\left(  -\frac{11\mp
3}{90\pi}\kappa^{2}-i\sqrt{\frac{3}{2}}\frac{3\mp1}{16}\kappa {\rm
e}^{-8/3\kappa} \right),
\label{Eq:small-kappa}
\end{equation}
$$
\kappa \ll1.
$$
The opposite asymptotics is
\begin{equation}
m^{2}_{\gamma ,\parallel(\perp)}
\approx
\alpha m_{e}^{2}
\frac{5\mp1}{28\pi^{2}}\sqrt{3}\Gamma^{4}(2/3)
(1-i\sqrt{3})(3\kappa)^{2/3},
\label{Eq:large-kappa}
\end{equation}
$$
1\ll \kappa \ll \alpha^{-3/2}.
$$
Note that at $\alpha \kappa^{2/3}\gtrsim 1$, the photon mass and electron
mass are of the same order and the approximation of Ref.~\cite{Ritus} does
not work.

We arrive to the expression
\[
\Delta_{Q,\parallel(\perp)}=1.49 \times 10^{13}~{\rm pc}^{-1}
\left(\frac{\omega}{10^{19}~{\rm eV}} \right)^{-1}
F_{\parallel(\perp)}(\kappa),
\]
where
\[
F_{\parallel(\perp)}(\kappa)
=
\frac{m^{2}_{\gamma ,\parallel(\perp)}(\kappa)}{\alpha m_{e}^{2}}
\]
is a function of $\kappa$ plotted in
Fig.~\ref{fig:ritus}.

For constant magnetic field and
electron density, the conversion probability is
%\begin{equation}
\[
P=\frac{4 \Delta _M^2}{\left(\Delta _p+\Delta_{Q,\perp}-\Delta_m
\right)^2+4 \Delta_M^2 } \sin^2\left( \frac{1}{2}{L \Delta_{\rm osc}}
\right),
\label{eq:conversion}
\]
%\end{equation}
where
\[
\Delta_{\rm osc}^2=\left(\Delta_p+\Delta_{Q,\perp}-\Delta_m \right)^2+4
\Delta_M^2
\]
and we assumed that imaginary parts of all $\Delta$'s can be neglected.
If $B$ and $n_e$ change spatially, the probability can be found by a
numerical solution of Eqns.~(\ref{schrodinger}).
The condition for the strong mixing is
\begin{equation}
4\Delta_M^2                                       \gg
\left(\Delta_p+\Delta_{Q,\perp}-\Delta_m \right)^2.
\label{Eq:strong-mixing}
\end{equation}
In an earlier version of this paper, we neglected the
contribution of $\Delta_{Q}$ and arrived at the conclusion that for
certain values of the parameters, conditions for strong photon-axion
mixing are satisfied in the blazar and in the Milky Way, but not in the
intergalactic space, both for very-high-energy (TeV) and ultra-high-energy
($10^{19}$~eV) gamma rays. However, as it has been pointed out e.g.\ in
Ref.~\cite{criticism}, using Eq. for the real part of the QED-induced
photon mass and given its negative sign, the condition
(\ref{Eq:strong-mixing}) can be satisfied only in the case
$$
\Delta_{Q,\perp}\ll \Delta_M
$$
which reads as
\begin{equation}
F_\perp(\kappa) \ll 2.33 \times 10^{-11} \kappa
\left(\frac{M}{10^{10}~\rm{GeV}} \right)^{-1}.
\label{Eq:kappa-condition}
\end{equation}
In Ref.~\cite{criticism}, the small-$\kappa$ expansion,
Eq.~(\ref{Eq:small-kappa}), was used, which results in the condition
\begin{equation}
\kappa \ll 3.31 \times 10^{-9}
\left(\frac{M}{10^{10}~\rm{GeV}} \right)^{-1}
\label{Eq:strong-mix-small-kappa}
\end{equation}
or equivalently
$$
\left(\frac{B}{\rm G} \right) \left(\frac{\omega}{10^{19}~{\rm eV}}
\right) \ll 1.07 \times 10^{-9}
\left(\frac{M}{10^{10}~\rm{GeV}} \right)^{-1}
.
$$
An alternative approach is to make use of the change of sign of
$F_\perp (\kappa)$ which was suggested in
Ref.~\cite{Chou}.  Neglecting the possibility of precise cancellations (to
many decimal points) between $\Delta_m$ and $\Delta_Q$, this means that
one should have $\kappa=\kappa_0 \approx 15$. However, in this case, the
imaginary part of the photon mass is much larger than $\Delta_M$ and
a photon produces an electron-positron pair much quickly than it is
converted to an axion.
%To have very large
%$\kappa$ and to exploit the fact that the LHS of
%Eq.~(\ref{Eq:kappa-condition}) grows as $\kappa^{2/3}$, see
%Eq.~(\ref{Eq:large-kappa}), while the RHS grows as $\kappa$, would require
%crasy $\kappa>10^{26}$. Recall that the calculation works for
%$\kappa<1600$.
It would be interesting to understand what happens in the
strong quantum regime $\kappa>\alpha^{-3/2}$ since for
$\omega=10^{19}$~eV, this regime corresponds to fields of
$10^4$~G which are not extremely large.  We see that the only possible  way to obtain strong mixing in
the weak-coupling regime is to satisfy
Eq.~(\ref{Eq:strong-mix-small-kappa}).

Other maximal-mixing
conditions, which also must be met, are
$$
\Delta_m \ll 2 \Delta_M,
$$
and
$$
\Delta_p \ll 2 \Delta_M,
$$
which are equivalent to
\begin{equation}
\omega \gg
255~\mbox{eV}
\left(\frac{m}{10^{-9}~\mbox{eV}} \right)^2
\left(\frac{B}{\mbox{G}} \right)^{-1}
\left(\frac{M}{10^{10}~\mbox{GeV}} \right),
\label{eq:max-mix-m}
\end{equation}
\begin{equation}
n_e \ll
2.8\times 10^{19}~\mbox{cm}^{-3}
\left(\frac{\omega }{10^{19}~\mbox{eV}} \right)
\left(\frac{B}{\mbox{G}} \right)
\left(\frac{M}{10^{10}~\mbox{GeV}}  \right)^{-1}.
\label{eq:max-mix-p}
\end{equation}
In addition, to have large mixing one should require that the size $L$ of
the region in which conditions (\ref{Eq:strong-mix-small-kappa}),
(\ref{eq:max-mix-m}), (\ref{eq:max-mix-p})
are fulfilled should exceed the oscillation length,
$$
L \gtrsim \frac{\pi}{\Delta_{\rm osc}},
$$
that is
\begin{equation}
L \gtrsim
0.01~\mbox{pc}
\left(\frac{B}{\mbox{G}} \right)^{-1}
\left(\frac{M}{10^{10}~\mbox{GeV}} \right).
\label{eq:max-mix-losc}
\end{equation}

\begin{figure}
\begin{center}
\includegraphics[width=0.9\columnwidth]{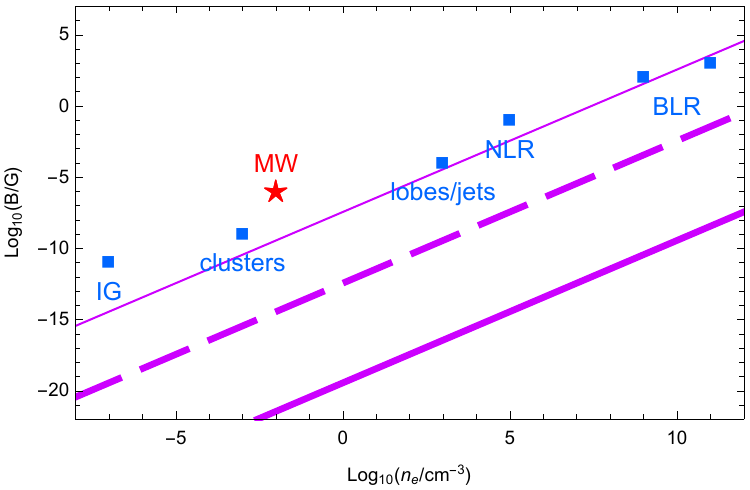}
\caption{
\label{fig:axi-n_e}
Typical values of the magnetic field~\cite{sources1} and electron density
\cite{CarrollOstlie, deAngelis0707.2695} in various astrophysical objects
(IG: intergalactic space, MW: the Milky Way, NLR and BLR: narrow- and
broad-line regions in active galactic nuclei). The condition
(\ref{eq:max-mix-p}) is satisfied above the thick line for energies
$\omega >10^{19}$~eV, above the dashed line for $\omega >1$~TeV and above
the thin line for $\omega >10$~MeV (for $M=10^{10}$~GeV). }
\end{center}
\end{figure}

\begin{figure}
\begin{center}
\includegraphics[width=0.9\columnwidth]{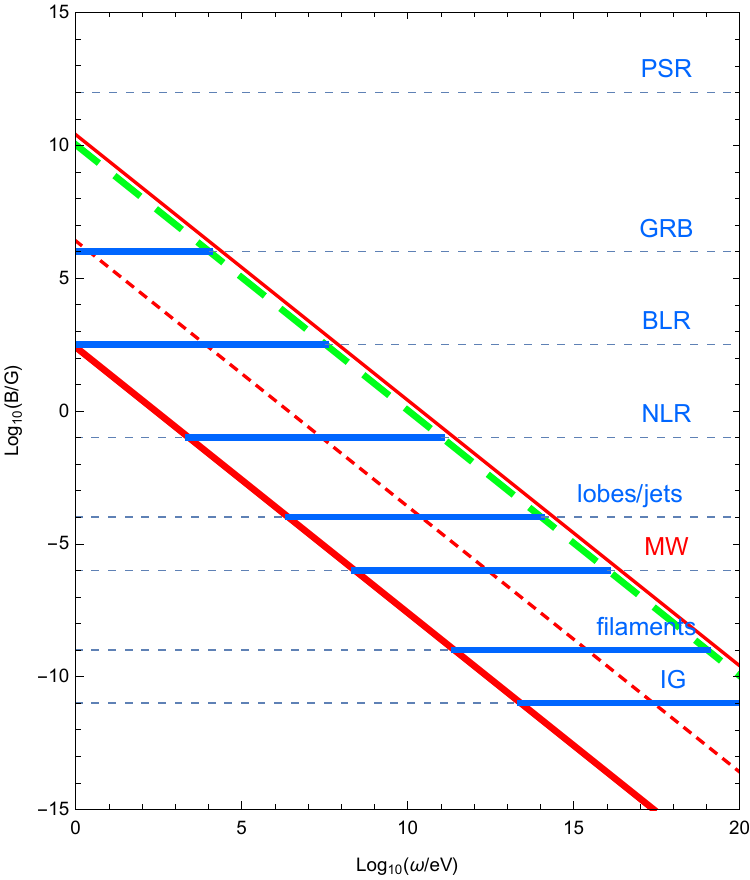}
\caption{
\label{fig:axi-m}
The conditions (\ref{Eq:strong-mix-small-kappa}) and (\ref{eq:max-mix-m})
on the parameter plane ``photon energy'' -- ``magnetic field''. The
condition (\ref{eq:max-mix-m}) is satisfied above red lines (thin for
$m=10^{-5}$~eV, dotted for $m=10^{-7}$~eV, thick for $m=10^{-9}$~eV).
Horizontal lines indicate typical $B$ values for various astrophysical
sources, the condition (\ref{Eq:strong-mix-small-kappa}) is satisfied
below a thick dashed green line: the mixing is possible below the green
line but above the red lines as indicated, for $m=10^{-9}$~eV, by the thick
blue parts of the horizontal lines
%cf.\ Fig.
%\ref{fig:axi-n_e}).
%Thick blue (yellow for the Milky Way) parts of the
%horizontal lines correspond to the energy range where the conditions
%(\ref{eq:max-mix-m}) are satisfied for
%$m=10^{-9}$~eV
(for $M=10^{10}$~GeV).
}
\end{center}
\end{figure}

From Fig.~\ref{fig:axi-n_e},
one sees that Eq.~(\ref{eq:max-mix-p}) is certainly fulfilled for
ultra-high-energy particles in all astrophysical gamma-ray sources. For
axion-photon coupling close to its experimental limit, the condition
(\ref{eq:max-mix-p}) is met down to energies as low as $\sim10$~MeV. The
conditions (\ref{Eq:strong-mix-small-kappa}) and
(\ref{eq:max-mix-m}) are illustrated in Fig.~\ref{fig:axi-m}.

\begin{figure}
\begin{center}
\includegraphics[width=0.9\columnwidth]{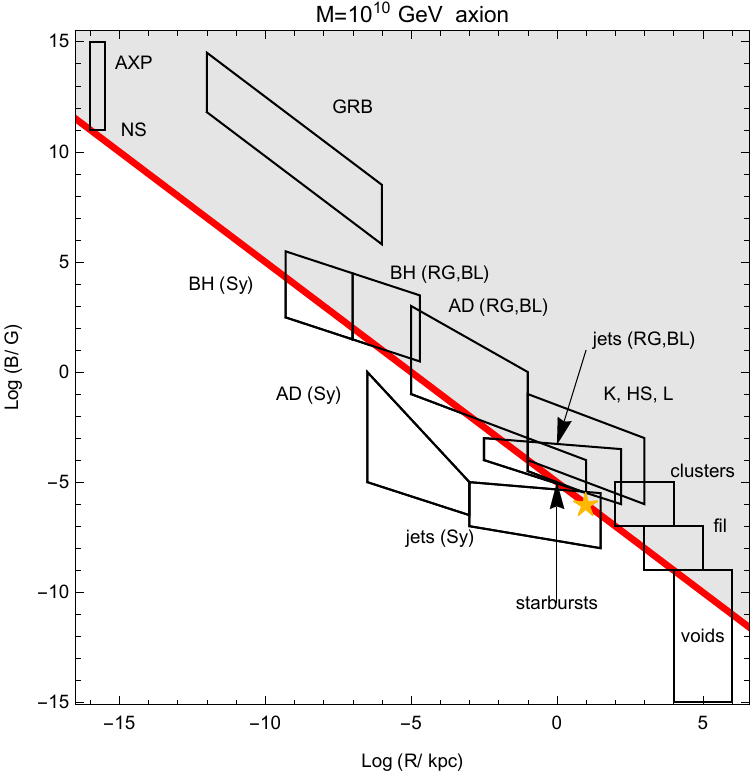}
\caption{
\label{fig:axi-Hillas}
The condition (\ref{eq:max-mix-losc}) for $M=10^{10}$~GeV on the updated
Hillas plot~\cite{sources1}. The condition is satisfied in the shadowed
region. The Milky Way parameters are denoted by a star. Also shown are
parameters for anomalous X-ray pulsars and magnetars (AXP), neutron
stars (NS), central black holes (BH) and for the central few parsecs (AD)
of active galaxies (low-power Seyfert galaxies (Sy), powerful radio
galaxies (RG) and blazars (BL)), relativistic jets, knots (K), hot spots
(HS) and lobes (L) of powerful active galaxies (RG and BL);
non-relativistic jets of low-power galaxies (Sy); starburst galaxies;
gamma-ray bursts (GRB); galaxy clusters and intercluster voids. }
\end{center}
\end{figure}

The condition (\ref{eq:max-mix-losc}), also very restrictive,
depends on both the size and the magnitude of the magnetic field and
can be superimposed \cite{FRT} on the Hillas plot for various
astrophysical sources (Fig.~\ref{fig:axi-Hillas}).
We see that if an axion like particle exists with the mass and coupling
outlined above, high-energy photons readily mix with it in many
astrophysical objects and environments. As a result, the axion flux
$F_a=F_\gamma/2$ accompanies the gamma-ray flux $F_\gamma $ independently
of the gamma-ray emission mechanism (for the maximal mixing, fluxes of
axions and of photons of each polarisation are equal).

\section{Neutral particles from distant sources}
\label{sec:BLL:axion:MW}

A number of studies suggest that a correlation may exist between the
arrival directions of cosmic rays and catalogues of BL Lac objects.  This
correlation exists without taking into account the magnetic field of the
galaxy, suggesting that the cosmic rays experience zero deflection as they
traverse this field and are therefore neutral particles, challenging
conventional models of cosmic-ray physics.

These claims are based upon two samples of cosmic rays, the first sample
combines events from the Akeno Giant Air-Shower Array (AGASA) of cosmic
rays with estimated primary energies $E>4.8 \times 10^{19}$~eV) and a
sample from the Yakutsk Extensive Air Shower Array (Yakutsk) of events
with estimate primary energy $E>2.4 \times 10^{19}$~eV. An excess of
correlations between the position of BL Lacs and the arrival direction of
cosmic rays in this combined data set was seen at separations less than
2.5$^\circ$ \cite{TT:BL}.

Similarly,  a sample of events with $E>10^{19}$~eV observed by the High
Resolution Fly's Eye detector (HiRes) tested positive for correlations
between source and BL Lac objects at angular separations less than
0.8$^\circ$ \cite{GTTT:HiRes}.

In both cases the separation was consistent with the detector's angular
resolution (which was much better in HiRes than in AGASA and Yakutsk). The
correlation with the HiRes sample was confirmed in an unbinned study and
was found to extend to lower energies \cite{HiRes:BL}.  The probability to
observe the correlation with three independent experiments by chance was
estimated by \cite{comparative} as $3\times 10^{-5}$ by a Monte-Carlo
study.

The correlation between BL Lacs and UHECRs seen in HiRes data
\cite{GTTT:HiRes, HiRes:BL} has been tested by the Pierre Auger
Collaboration \cite{Auger:BL} and no positive signal has yet been found.
However, it turns out that this is unsurprising for the following three
reasons: --

Firstly, Pierre Auger is located in the Southern hemisphere and sees
different BL Lacs to other experiments and due to incompleteness of the
astronomical catalogs, fewer potential UHECR emitters are known in the
South.

Secondly, the angular resolution of the Pierre Auger array is also inferior
to that of the HiRes stereoscopic telescope which means that the
sensitivity to such correlations can only be achieved with much more data.

Finally, as has been pointed out in \cite{GTTT:HiRes} and \cite{HiRes:BL}
and further discussed in \cite{TT:neutral}, the correlation observed by
HiRes implies neutral cosmic particles traveling for cosmological
distances, a fact which requires unconventional physics. Most probably the
primary particles of the resulting air showers are neither protons nor
nuclei. However, the energy determination of the PA surface detector is
extremely sensitive to the type of the primary cosmic particle because of
very strong sensitivity of water tanks to muons in the air shower. For
instance, energies of gamma rays are always underestimated by a factor of
a few (see e.g.\ \cite{Billoir, Nuhuil}). Due to the steeply falling
spectrum of UHECRs, this may dilute the observed signal.  It would
therefore be interesting if the Pierre Auger collaboration were able
search for the correlation using their Fluorescence detectors rather than
the water tanks.

An independent test of the cosmic ray -- BL Lac correlation is underway
\cite{Igor-TA} with the Telescope Array experiment located in the Northern
hemisphere and equipped with the array of scintillator detectors and
fluorescent telescopes capable of stereo imaging.

Having discussed the evidence for the correlation between the arrival
direction of ultra high energy cosmic rays and BL Lac objects, we will
move on to look at the use of axions to explain how neutral particles
could traverse the Universe without complete attenuation.

\section{Axions as ultra high energy cosmic rays.}
\label{expl}
In the framework of the Standard model of particle physics and assuming
standard astrophysics, neutral particles with energies $\sim 10^{18}$~eV
cannot propagate for $ \gtrsim 100$~Mpc, the distance to the nearest BL
Lacs.  The only exception is neutrino which can be excluded as an
explanation for these events by considering the height of development of
the atmospheric showers and noting that they are not close enough to the
ground to be consistent with the weak interaction cross sections.

\begin{figure}
\begin{center}
\includegraphics[width=0.9\columnwidth]{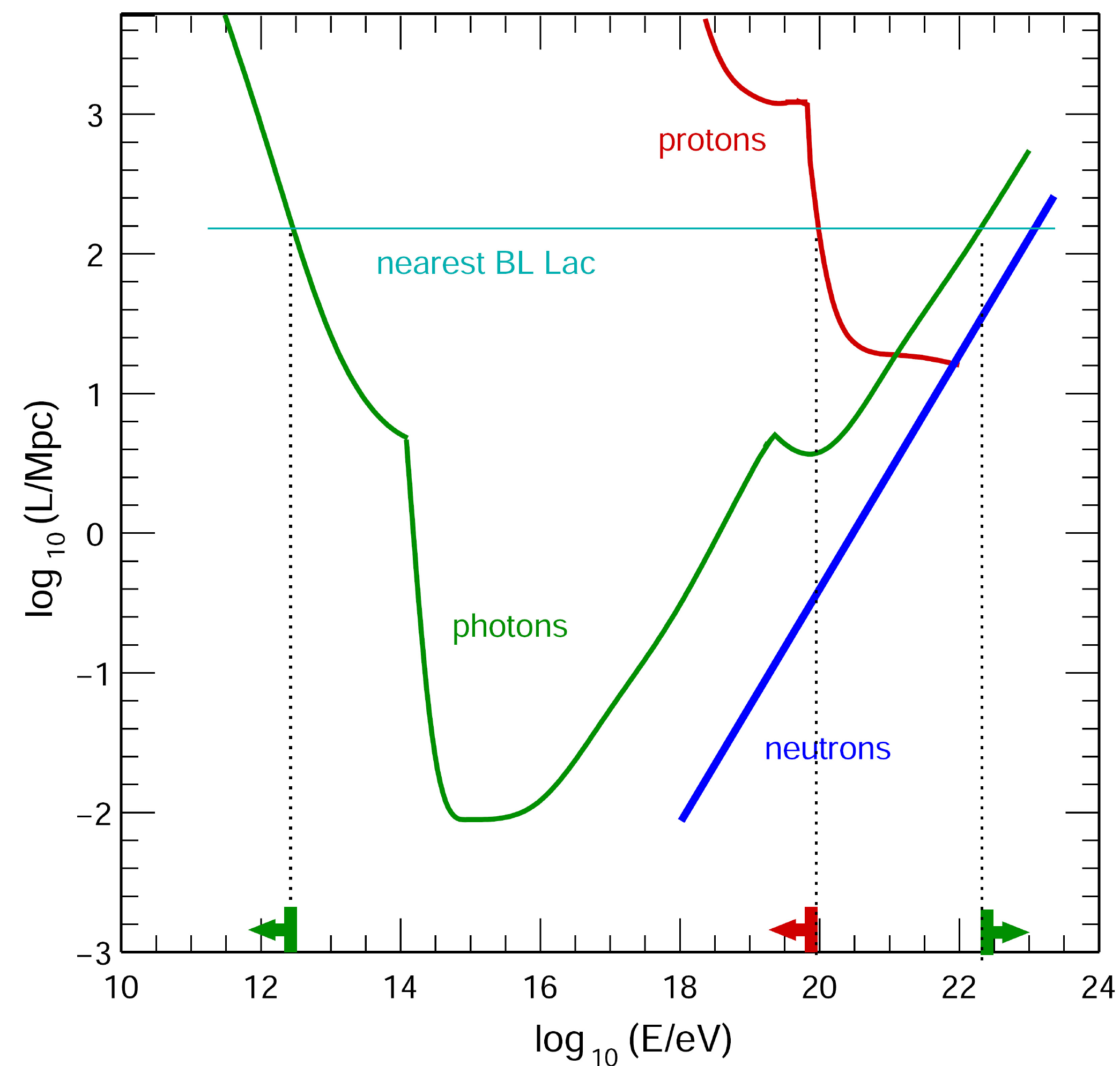}~~~~~~~
\caption{
\label{fig:attenuation}
Attenuation length of different kinds of particles
as compared to the distance to the nearest BL Lac (blue horizontal line)
and the size of the Universe (upper bound of the plot). The green line
corresponds to photons, blue to neutrons and red (shown for comparison) to
protons. The lines for protons and photons are taken from the review
\cite{Boratav}.}
\end{center}
\end{figure}

Photons interact with the background radiation which results in pair
production and the development of electromagnetic cascades.  Known
unstable particles decay at much shorter distances
(Fig.~\ref{fig:attenuation}).
In the framework of more involved descriptions which do not require new
physics, the neutral particles may be created in interactions of protons
inside or not far from the Milky Way; however, in this case the observed
effect also cannot be explained \cite{TT:neutral}.

Even beyond the Standard model it is difficult to find a non-contradictory
explanation of the observed correlations. New stable strongly interacting
particles \cite{Farrar:gluino, Semikoz:b-squark} should be heavy enough
not to be detected in accelerators, but the probability to create such a
heavy particle is low and therefore one would expect for each one of these
particles that there should be huge fluxes of accompanying radiation in
conflict with constraints on diffuse cosmic gamma radiation.

Models which suggest the existence of a relatively heavy ($\sim$MeV) axion
like particle (sgoldstino) \cite{sgoldstino} suffer the same problem.

In the models where there is an enhanced neutrino-air cross section (see
e.g.\ \cite{nu-air}), besides some theoretical difficulties, the cross
section rise is not sufficient to explain the shower development. Only in
the models with Lorentz-invariance violation \cite{Lorentz-violation}
decaying neutral particles (neutron or $\pi^0$ meson) might be stable in a
certain energy region and propagate to cosmological distances.  It can be argued however that postulating the existence of a new particle is less
drastic than altering the framework of relativity.

The scenario which we investigate here is based on the mixing of photon
with light axions. The parameters of the model which work in explaining
the conundrum of the neutral primaries outlined above do not contradict
any experimental limits and may allow one to explain some other
astrophysical puzzles as well.

The maximal mixing conditions (\ref{Eq:strong-mix-small-kappa}),
(\ref{eq:max-mix-m}), (\ref{eq:max-mix-p}), (\ref{eq:max-mix-losc}) are
satisfied (cf.\ Figs.~\ref{fig:axi-n_e}, \ref{fig:axi-m},
\ref{fig:axi-Hillas}) for various astrophysical objects, allowing for
different scenarios of axion-photon transitions which might be relevant
for the BL Lac correlation. We summarize them in
Table~\ref{table:scenarios} and describe in more detail below. For
convenience, we include also the information about TeV photon mixing
relevant for the gamma-ray observations. As we will see, the choice of a
particular scenario depends on the value of the intergalactic magnetic
field (IGMF) at scales $\gtrsim$~Mpc which at present is poorly known.
\begin{table}
\begin{center}
\begin{tabular}{ccccccccc}
\hline
No. & $m$ & IGMF & $\omega$ & \multicolumn{4}{c}{strong mixing
in} & dominant\\
\cline{5-8}
    &  eV &  G   &  eV      & BL & fil & IG & MW &
conversion     \\
\hline\hline
1 & $\sim 10^{-7}$& $\lesssim 10^{-11}$&$ 10^{12}$&+&--&--&+&source+MW\\
  &               &                    &$ 10^{19}$&--&+&--&--&fil+fil\\
2 & $\sim 10^{-7}$& $\sim 10^{-9}$&$ 10^{12}$&+&--&--&+&source+MW\\
&               &                    &$ 10^{19}$&--&+&+&--&IGMF+IGMF\\
3 & $\sim 10^{-5}$& any&$ 10^{12}$&+&--&--&--&no explanation\\
&               &                    &$ 10^{19}$&--&+&--&--&fil+fil\\
&               &                    &&&&&&(IGMF if strong)\\
4 & $\lesssim 10^{-9}$& $\sim 10^{-9}$&$ 10^{12}$&+&+&+&+&IGMF+IGMF\\
&               &                    &$ 10^{19}$&--&--&+&--&IGMF+IGMF\\
\hline
\end{tabular}
\end{center}
\caption{\label{table:scenarios}Different scenarios for mixing of
high-energy cosmic photons with axions. Columns give: the number of the
scenario (as referred to in the main text); the axion mass $m$; the assumed
value of IGMF in large-scale voids; the energy of photons $\omega$ (two
cases are presented, $\omega\sim 10^{12}$~eV relevant for TeV gamma rays
from distant blazars and $\omega\sim 10^{19}$~eV relevant for the cosmic
ray-- BL Lac correlations); potential sites where the strong mixing is
possible (+) or not (--); and the principal sites of the $\gamma \to a$
and $a\to \gamma$ conversion (BL=BL Lacs, fil=filaments, IG=intergalactic
voids, MW=Milky Way). More details are discussed in the text.}
\end{table}

\underline{Case~1: $m\sim 10^{-7}$~eV, weak IGMF.} From
Fig.~\ref{fig:axi-m}, it is clear that conditions
(\ref{Eq:strong-mix-small-kappa}), (\ref{eq:max-mix-m}) leave a window of
$\sim(10^{-13}\dots 10^{-9})$~G for conversion of $\sim 10^{19}$~eV
photons. If IGMF in voids is $\sim 10^{-11}$~G or weaker, conversion on it
is suppressed since the condition (\ref{eq:max-mix-m}) is not satisfied.
Intense photon-axion conversion may happen in the regions of a few
Megaparsec size with the magnetic field $\sim 10^{-9}$~G. According to
simulations of Ref.~\cite{DolagMF}, these conditions are satisfied in
certain elements of the large-scale structure of the Universe which we
somewhat loosely call ``filaments'' for brevity. In this case,
protons are accelerated to ultra-high ($E \gtrsim
10^{20}$~eV) energies in the sources (according to Ref.~\cite{sources1},
acceleration of protons in BL Lacs up to these energies contradicts
neither the Hillas criterion nor the radiation losses). Interaction of
these protons with the intense blazar emission results in the pion
photo-production similar to the GZK effect which for a fraction of the
accelerated particles takes place directly in the source. If the source is
located in, or near, a ``filament'', then intensive mixing there
converts 1/3 of photons into the axions of the same energy
so the axion-photon beam propagates into space towards earth.
Further mixing in intergalactic space before the photon-axion beam arrives
at the local ``filament'' where the observer sits is suppressed due to
small magnetic fields in voids. The photon part of the beam
interacts with background photons and loses energy while the axion part
propagates unattenuated. Then, upon arrival at the local ``filament'' where
the magnetic field is several orders of magnitude higher than in voids,
intensive mixing again takes place and a significant
fraction (2/3 for maximal mixing) of the axions are converted back into
photons which are then detected as neutral particles from BL Lacs.  The
maximum fraction of photons detected in cosmic ray detectors on earth can
be 2/9 of the total flux of photons of same energy emitted in the source.
We note that for the parameters of this case, the mixing in IGMF is not
possible for $\omega\sim 10^{12}$~eV either; nor mixing is possible in the
``filaments'' for these energies. However, the scenario of
Ref.~\cite{serpico} (mixing in the source and in the Milky Way) works for
TeV photons.

\underline{Case~2: $m\sim 10^{-7}$~eV, strong IGMF.} The condition
(\ref{eq:max-mix-losc}) is satisfied for IGMF $\sim 10^{-9}$~G, so the
dominant place of conversion of UHE photons is IGMF in this case. At the
same time, for $\omega \sim 10^{12}$~eV, the condition
(\ref{eq:max-mix-m}) forbids strong mixing at IGMF and the ``source--Milky
Way'' mechanism is again operational for TeV photons.

\underline{Case~3: $m\sim 10^{-5}$~eV.} In this case, the conditions
(\ref{Eq:strong-mix-small-kappa}) and (\ref{eq:max-mix-m}) leave a very
narrow strip on the ``$\omega-B$'' parameter plane, Fig.~\ref{fig:axi-m}.
For UHE photons, this strip allows for the conversion at $B\sim
10^{-9}$~G, that is in ``filaments'' for weak IGMF and in voids for the
strong one. No viable conversion scenario exists for TeV photons in this
case.

\underline{Case~4: $m\lesssim 10^{-9}$~eV, strong IGMF.} This is a
realization of the scenario of Refs.~\cite{roncadelli, DARMA2011} of
conversion at IGMF, working for both $\omega \sim 10^{12}$~eV and $\omega
\sim 10^{19}$~eV. For TeV photons, other conversion sites are possible but
their effect is negligible compared to that of long-distance IGMF.

We see that the applicability of various scenarios is strongly dependent
of the assumed values of IGMF. Current observational limits (see e.g.\
Ref.~\cite{IGMFlimits} for a review) constrain the magnetic fields at
$\gtrsim $Mpc scale to be in the range $(10^{-16}\dots 10^{-9})$~G, the
lower bound\footnote{This bound may change in the presence of axions.}
coming from non-observation of GeV emission from certain TeV sources
\cite{IGMFlow-lim} while the upper one coming from the CMB polarization
\cite{IGMF-CMBpolariz} and Faraday rotation measurements
\cite{IGMF-Faraday}. The simulations of Ref.~\cite{DolagMF} favour very
low ($\sim 10^{-15}$~G) magnetic fields in the large-scale voids
(otherwise far too high fields in the galaxy clusters are produced,
incompatible with observations). At the same time, $\sim 10^{-9}$~G fields
are obtained in this simulations for certain few-megaparsec scale parts of
the filaments. There are also other indications to very weak magnetic
fields in the voids \cite{IGMF-Semikoz} but these are model-dependent. The
CMB measurements by the Planck satellite, currently in flight, will test
the IGMF in the range $(\sim 10^{-11}\dots 10^{-9})$~G, crucial for the
choice of the axion conversion scenario.

In the case of conversion in voids, that is of strong ($\sim 10^{-9}$~G)
IGMF, one cannot use directly the oscillation formalism outlined above for
UHE gamma rays and axions because the attenuation length of $\omega\sim
10^{19}$~eV photons on the radio background radiation is $l\sim3$~Mpc,
cf.\ Fig.~\ref{fig:attenuation} (the precise value of $l$ is sensitive
to the poorly known intergalactic radio background). Within our precision
and given lack of knowledge of the field, we may estimate the flux of the
photons arriving to the Earth as a $P(l)^{2}\sim \left(l/D \right)^{2}\sim
(10^{-3}\dots 10^{-4})$ fraction of the initial photon number flux, where
$D$ is the distance to the source. This estimate does not contradict the observed UHE cosmic-ray flux within the photoproduction scenario for
UHE gamma rays. The remaining part of the flux interacts with the cosmic
background radiation and experiences electromagnetic cascades down to
photons of $\sim$GeV energies for whom the Universe is transparent.
Depending on the radiation and magnetic field strengths in the source, in
its close environment and along the trajectory to the Earth, these
secondary GeV photons contribute either to the (extended) image of the
source or to the diffuse gamma-ray background. The corresponding flux may
be estimated as follows. The flux of events correlated with BL Lacs
is~\cite{predictions} roughly 0.03 of the total cosmic-ray flux at
$10^{19}$~eV. The latter flux as detected by HiRes~\cite{HiRes-flux} is
$J_{\rm CR}E^{3}\approx 2\times
10^{24}$~eV$^{2}$m$^{-2}$s$^{-1}$sr$^{-1}$. The estimate of the
corresponding flux of $E_{0}\sim$GeV photons may be obtained from the
energy conservation and in the IGMF-conversion scenraio reads as
\begin{equation}
J_{\rm IG} E_{0}\sim \left(\frac{D}{l}   \right)^{2} \, 0.03\, J_{\rm
CR}\frac{E^{2}}{E_{0}}\sim 10^{-6 } {\rm cm}^{-2} {\rm s}^{-1} {\rm
sr}^{-1}.
\label{Eq:C4*}
\end{equation}
This is of the same order as the diffuse GeV flux observed by Fermi
\cite{FermiDiffuse}. Given all uncertainties in our estimate, as well as
in the observational value \cite{FermiDiffuse, EGRETdiffuse1,
EGRETdiffuse2}
of the GeV flux, we do not use this number to constrain the scenario; with
the present precision it may, depending on the assumptions, either explain
the part of the GeV background unaccounted for by known contributors, or
overshoot the observed value thus indicating that the scenario is not
viable. If the magnetic fields and radiation backgrounds allow for
formation of an extended image of the source, then the flux in
Eq.(\ref{Eq:C4*}) should be distributed among the observed sources rather
than spread uniformly over $4\pi$~sr. The number of sources $N_{s}$ may be
estimated from the statistics of clustering~\cite{DubTin} as $N_{s}\sim
60$. In this case, the value of the single-source flux,
\[
J_{1,\rm IG}\sim \frac{4\pi~\rm sr }{N_{s}}J_{\rm IG}E_{0} \sim 10^{-7}
{\rm cm}^{-2} {\rm s}^{-1},
\]
is too high to be realistic for 60 sources.

On the other hand, for weak IGMF scenarios roughly 1/3 of original UHE
photons are converted to axions within a few Mpc from the source and 2/3
of them convert back to photons within a few Mpc from the Earth.
Therefore, instead of $(l/D)^{2}$, the observed flux constitutes $\sim
2/9$ of the emitted one. The flux of a single source is then
\[
J_{1,\rm F}\sim \frac{4\pi~\rm sr }{N_{s}} \, \frac{2}{9} \, 0.03\,
J_{\rm CR} \frac{E^{2}}{E_{0}}\sim 10^{-9}
{\rm cm}^{-2} {\rm s}^{-1},
\]
well within the Fermi sensitivity.
The angular size of
the halo is then
$
\theta\approx \frac{l}{D},
$
which is a fraction of a degree, well below the width of the point spread
function of either EGRET or Fermi. This means that the correlated BL Lacs
should be gamma-ray sources in agreement with the results of
Refs.~\cite{GTTT:EGRET, index}. A method of detection of the size of
extended images
of this kind is discussed in Ref.~\cite{occultation}.

\begin{figure}
\begin{center}
\includegraphics[width=0.9\columnwidth]{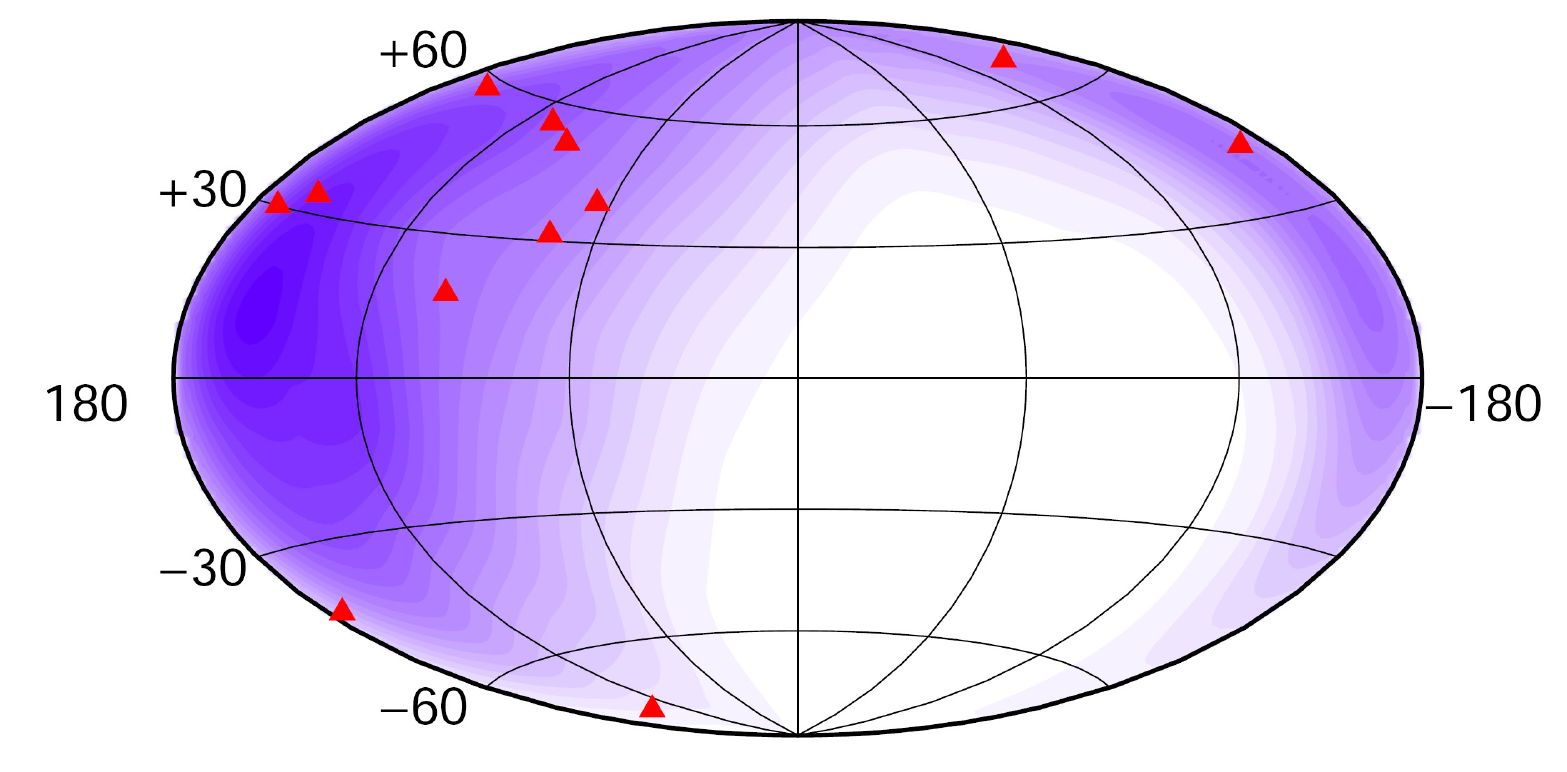}
\end{center}
\caption{\label{compare}
The exposure of the HiRes
cosmic ray detector - darker shaded regions have had more exposure.  The
red triangle data points correspond to cosmic rays correlated with the
position of BL Lacs. }
\end{figure}

Because the cosmic magnetic fields, in particular in ``filaments'', are
non-uniform, we expect the probability for axions which
arrive from far away to convert back into photons to depend upon the
direction they move before they arrive
at Earth.  This should reflect itself in the distribution of the arrival
directions of the correlated events. To look for the possible anisotropy,
we use the list of correlated BL Lacs from \cite{GTTT:HiRes} and
compare their distribution with respect to the experimental exposure
with the full sample of 156 BL Lacs studied in
\cite{GTTT:HiRes, HiRes:BL}. The correlated events are plotted,
together with the exposure, in Fig. \ref{compare}.
One sees that they do
not appear to follow exactly the expected random distribution (they would
be more likely to turn up in the most densely shaded region if the
distribution was isotropic).
To quantify these suspicions, we use the method recently becoming popular in
tests of global anisotropy of UHECR arrival directions
\cite{AugerComment, WaxmanKashti, TinyakovKoers}. For each BL Lac
with coordinates $(l_i,b_i)$ we calculate
the value of the experimental exposure towards this point of the sky,
$A_i=A(l_i,b_i)$. Then we compare, by means of the Kolmogorov-Smirnov
test, the distributions of these $A_i$ for BL Lacs which are correlated
with cosmic rays and for all BL Lacs, correlated or
not\footnote{We use the full BL Lac sample instead of simulated isotropic
samples to account for non-uniform distribution of BL Lacs in the catalog.
}. The test
gives a probability of 0.024 that the two distributions of $A_i$ are
realizations of the same distribution, thus disfavoring the idea that
correlated events come from random/isotropic regions on the sky.  Though
it is not possible to judge, without a quantitative model of magnetic
fields outside the Galaxy, whether the non-uniformity is related to the
field structure at the Megaparsec scale, it is tempting to note
that a similar deviation from isotropy would be expected in our weak-IGMF
scenarios (cases 1 and 3).

%%%%%%%%%%%%%%%%%%%%%%%%%%%%%%%%%%%%%%%%%%%%%%%%%%%%%%%%%%%%%%%%%%%%%%%%%%%%%%%%%%%%%%%%%%%%%%%%%%

\section{Discussion and conclusions}
\label{sec:conseq}
We have shown in the previous section that there is some motivation for a
possible interpretation of the neutral events correlated with the position
of BL Lac objects in the sky being due to photons that have been able to
traverse the Universe because of their conversion into axions
and then back into photons.  More data with regards to the intergalactic
magnetic field, especially at the Megaparsec scale, and more cosmic ray
events will be able to add or subtract confidence in this interpretation
but the scenario leads to several other consequences which may be tested
in future studies.

{\it Primary particle type of the correlated events.}
Clearly, if axions are the explanation, then the primary particles of the
correlated events should be photons. Currently studies of the primary
particle type for the HiRes events are not published. The photon-primary
hypothesis agrees perfectly with the absence of correlations in the Auger
surface-detector data~\cite{Auger:BL}: the photon energies are
underestimated \cite{Billoir} by this detector by a factor of four on
average \cite{Nuhuil}, so that the correlated events would be lost among a
large number of hadronic events of lower energy.  Such a situation should
also be the case in the future data, although as alluded to earlier, the
correlations should be seen in the data of fluorescent detectors of Pierre
Auger and Telescope Array and in the surface detector of Telescope Array.

{\it Secondary photons and the extended image.}
As discussed above in Sec.~\ref{expl}, in certain scenarios the extended
image of the source in GeV photons is formed and may be detected. Other
scenarios result in a contribution to the GeV diffuse background. Both
possibilities may be constrained with the Fermi data.

{\it Axion parameters.} The model requires the axion-like particle with
mass
$m \lesssim 10^{-5}$~eV and the inverse coupling to photon close to the
current experimental limits,
$M\sim (1\div 10) \times 10^{10}$~GeV.
The most direct confirmation of the scenario would come from the
discovery of that particle. This region of the parameter space is
available for exploration with CAST at sufficiently large exposure. The
axion with these parameters may also affect the polarization of
extragalactic radio sources
\cite{Gnedin:axion, polarisGRB, Polaris:axion:new}.

To summarize, the existence of an axion-like particle with an inverse
coupling $M\sim 10^{10}$~GeV and a low mass $m\lesssim 10^{-5}$~eV has been
invoked by other authors to explain the detection on earth of TeV photons
from cosmological sources - flux which is difficult to explain given that
such photons should produce electron-positron pairs on the cosmic infrared
background \cite{roncadelli, serpico}.  The presence of such a particle
would enable some photons to convert into axions and to travel
over the intervening space without interacting with the background
radiation before turning back into photons.

In this work we have tried to use the same method to explain a set of
ultra high energy cosmic ray events which seem to come from BL Lac
objects.  Since such events seem to lead straight back to the source, the
particles should be neutral because charged particles would be deflected
by the magnetic field of the galaxy.  However, no neutral particles in the
standard model seem capable of traversing the Universe at such a high
energy.  The same idea of photons turning into axions and
then back into photons is immediately applicable to this
cosmic ray situation.

Clearly, more data is needed to show whether or not this correlation has
occurred by chance, both cosmic ray data and data on the arrival of TeV
photons from cosmological sources will help to add support to or rule out
this hypothesis.

It has been suggested that the existence of a light  axion-like particle
would also help explain some other astrophysical conundrums such as the
white dwarf luminosity function~\cite{wd}.

Finally it is appropriate to re-iterate that the parameters of interest
for this effect suggest a weak coupling for the axion, but not so weak
that it cannot be probed by experiments such as CAST~\cite{CAST2008} or
the new generation axion helioscope~\cite{NGAH}. The low mass required to
ensure that one is in the region of maximal mixing means that such an
axion should be able to be ruled out or confirmed by one of these
experiments.

We are grateful for discussions with
O.~Kalashev, G.~Rubtsov, V.~Rubakov, D.~Semikoz, P.~Tinyakov and
I.~Tkachev. We thank D.~Krotov and G.~Rubtsov for their help in obtaining
 the full text of Ref.~\cite{Toll} and N.~Mikheev for bringing
 Ref.~\cite{Ritus} to our attention. This work was supported in part by
the
 grants RFBR 10-02-01406, 11-02-01528 and FASI 02.740.11.0244, by the
 grant of the President of the Russian Federation NS-5525.2010.2 and by
 the Dynasty foundation (ST).

\end{document}